\documentclass{ws-p8-50x6-00}

\begin{document}

\title{A Closing Talk For A Very Nice Meeting}

\author{Bo Andersson}

\address{Department of Theoretical Physics
University of Lund, Lund Sweden\\ 
E-mail: bo@thep.lu.se}


\maketitle

\abstracts{
I will use this opportunity on the one hand to comment 
upon some of the many interesting results that have been presented at 
this meeting, on the other hand to discuss some new features that 
we have recently learned  on the structure and properties of the 
Lund Model both with respect to the fragmentation of 
multigluon string states and the partonic 
cascades based upon perturbative QCD.}

\section{Introduction}

I have been very much impressed about the quality and the amount of 
new results that have been presented at this meeting and I would 
like to start by applauding both the organisers and the speakers 
for excellent arrangements, entertainment and new insights.

It is evident that there is no way to comment upon all the results 
we have heard of during the five days and consequently I will 
be satisfied to take up a few of them and to briefly consider
their significance. I apologize to all those that I am not going to 
mention.

After that I will in the spirit of a Closing Talk go over to those 
parts of the multiparticle dynamics that are close to my own heart. 
After a brief lamentation on our present way to treat QCD I will start 
by pointing out that there are strong indications of the existence of 
a dynamical scale between one and two GeV that ought to be taken into 
account in all considerations of a confined field theory. 

After that I will 
briefly touch upn the new features that we have found in the 
structure of Lund String Fragmentation both with and without 
multigluon excitations. There is also another (and even more recent) 
find dealing with  the properties of the QCD perturbative cascades. 

One of the 
major differences between the abelian QED and the non-abelian QCD is 
the fact that the QCD field quanta, the gluons, are charged and that 
consequently  emission of a gluon implies that the currents are changed. 
The phase space in QED for multiple emission of photons  is 
given by the properties of the original current (besides the effects 
of the recoils in the emission of ``hard'' photons). 

But already the 
emission of a first QCD  gluon in $e^+e^-$-annihilation means that the  
original $(q \bar{q})$ dipole is changed. As it happens the change 
is (to a very good approximation) from one dipole to 
two (independent) dipoles, one between the quark ($q$) and the 
first gluon and the other between the gluon and the anti-quark ($\bar{q}$).
While the first dipole is at rest in the total cms the two ``new'' 
dipoles are moving with respect to each other. This implies that the 
combined phase space for emitting either from one or the other dipole 
(this is what ``independence'' means) will cover a larger phase space region. 
In rapidity space the increase can be described as an extra region of a size
corresponding to the logarithm of the squared transverse momentum of 
the first gluon emission.

It is possible to describe the phase space from multiple gluon emission 
in terms of a generalised rapidity, $\lambda$ 
or as the length of a curve composed of connected 
hyperbolas related to the generalised rapidity in the same way as the 
length of a single hyperbola is related to ordinary rapidity. This curve 
is defined in an infrared stable way from the partonic energy momenta and 
it has (multi-)fractal properties with dimensions given by the so-called
anomalous dimensions of QCD.

The basic new result is that while the bremsstrahlung spectrum in QED 
can be described as a constant density (given by $\alpha/\pi$ with 
$\alpha$ the finestructure constant) in rapidity $y$ and the logarithm of the 
(squared) transverse momentum, $\kappa$, the QCD spectrum can 
be described {\em as a similar  constant density
in terms of the the generalised rapidity $\lambda$ and  $\kappa$}. 
I will discuss the implications of this result, thereby ending with the 
usual optimistic credo of a theoretical multiparticle physicist: 
``There is still a lot of interesting things to be found in phase space!''
(at least if you chose the right space).

\section{Some Remarks On What We Have Learned}
We firstly learned about the running of the chinese accelerator BES from Dr
Xu Guofa and we are happy about their precise results on the ratio 
between hadron- and muon production from $e^+e^-$-annihilation almost 
down to the treshold. We know that this work has a significance for the 
basic parameters of the Standard Model. Personally I feel that  
precise results from the
baryon-antibaryon channels should provide very useful constraints 
on all the multiparticle production models. As I understand it from 
discussions with Marek Karliner there are strong indications that the 
isospin one channel is dominating close to treshold, which would imply that 
the particle content in these channels is not given by ``ordinary'' 
quark counting. Let me also say that the energy region covered by BES 
is interesting in the sense that it covers a good deal of the 
mass spectrum of the ``clusters'' used in HERWIG. 

We also learned about the experiments on the CP-break from Dr Anzivino. 
I remember the times when we knew that there was a mixing parameter 
$\epsilon$ (known to be of the order of $10^{-3}$ from the Fitch-Cronin 
experiment) that made a connection between the  $K^0$ and $\bar{K^0}$ 
states possible. We also knew that there might be another  
CP breaking parameter $\epsilon^{\prime}$ connecting the neutral kaons 
to the two and three pion decay channels. This parameter may be vanishing 
or at least much smaller and I do not know the cosmological implications
of the beautiful experiments done now (NA48 and KTeV) with 
an $\epsilon^{\prime}$ of the order $10^{-6}$ . What is so impressing for
me is that the resulting experimental numbers again are converging quickly 
just as they evidently are for the $g-2$ experiment done over at Brookhaven.   
I am just waiting for the day when these precission experiments are going to
finally break our peace!

Pedro Abreu discussed a longstanding question, i.e. whether there are color 
(re- or inter-)connections in QCD. We know of one case where our 
present theory indicates the existence of such a mechanism, i.e. 
the decay of a $B^0$ into a $J/\Psi$ plus anything. The production mechanism 
is described in terms of the decay of the $\bar{b}$-quark
into a (color singlet) $W$ and $\bar{c}$-quark and the subsequent 
decay of the $W$ into among other a
$c$, that joins with the $\bar{c}$. It is evident that the $W$ is very, 
very far off-shell but it is nevertheless a fact that 
the size of the mechanism matches the experimental findings.  

I remember 
that Peter Zerwas came to Lund about twelve years 
ago and that he made  a  model with Gosta Gustafson and one of our 
students on the possible repercussions in the measurement of the 
$WW$-production in LEPII. I was scared by the very large effects that 
may occur if the  $W$'s decay into two $(q_j\bar{q}_j)$, $j=1,2$ and 
the partons afterwards rearrange themselves with ``the wrong partner''.  (As
soon as LEPII started we knew that this was not the case). 
In a  paper (before LEPII), Sj\"ostrand and Khoze, in very great 
detail showed that 
unless we would move far away from the $W$-poles only non-perturbative 
possibilities could produce effects. They went on to calculate such 
effects in a semi-classical model with overlaps between the decaying 
string fields. Their results was that even with a thirty percent overlap 
and subsequent color reconnection these nonperturbative  effects 
would only be of the order 
of $10-20$ $MeV$ on the $W$ width, i.e. ``below danger''. There have been 
further tries (L\"onnblad, even partly before S.-K., and 
Gustafson and H\"akkinen)
but it seems that if there is an effect it is terribly difficult to find.

Ingelman and his collaborators have introduced color reconnections to explain
``rapidity gap'' events in Deep Inelastic Scattering, noting that 
``the wrong colors'' may occur as $1/N_c^2 \simeq 10 \%$ corrections 
(which are in the right ball park for the observations). I am myself 
worried about the possibilities because a large part of the structure 
would go out of our present models if the fields are allowed to change 
color all over the place. I have personally taken recourse to the 
following argument. 

We should remember that in QED the fields are 
given almost exclusively by a knowledge of the charges but this is no 
longer so for the confined QCD. In the Lund Model we have always taken 
the QCD fields as the primary objects.
In every order of perturbation theory (as it is 
done in present day perturbative QCD) you may calculate the emitted 
charges  and then obtain interference between the different color charge 
configurations. But if we were able to sum up the whole result {\em then 
we may find superselection rules} saying that {\em the different color
field configurations do not interfere}.

Franz Mandl discussed his work with Brigitte Buschbeck on the properties of 
gluon jets. Their problem is again of a longstanding nature: do the 
gluon jets show the same particle content as the $q$ or $\bar{q}$ jets?
The Lund gluon model, where the gluons are internal excitations acting on 
the string fields with a force twice that of the $q$ and $\bar{q}$ endpoints 
is actually an extreme variety of all possible gluon models. Montvay showed 
in a Physics Letter from 1979 that you may build models in which the 
force from the gluon could be anything from twice down to zero times 
the endpoint force. Then the gluon would not be able to ``drag along'' 
the force field  all the way when it moves out. Instead there would be 
a kind of ``polyp'' dragged out behind it linking it to the field. When 
the PETRA-PEP machines started we learned that if there was such an effect 
then it would be small, i.e. the gluon force ought to be at least $180 \%$ 
of the endpoint force. Peterson and Walsh tried to make models in these 
days with production of ``glueballs'' and as I learned from Wolfgang Ochs 
he and Minkowski are still pursuing this question. We know that this 
is a very difficult subject (although there are persistent rumours that 
there may be something on the level of $1.5$ $GeV$). It is nevertheless 
evident that Buschbeck-Mandl (who are some of the most careful people I 
know of) seem to find something fishy at the end of the gluon jets. 
There is another really careful man, Bill Gary, who talked later about 
the multiplicities in gluon and $(q \bar{q})$ jets, and confirmed that 
you may hardly ever come up to the famous ratio of $9/4$. Maybe Bill 
can be coaxed to look also?

I would also like to briefly discuss the spin effects, which were considered 
in the talk by Dr Liang. Once upon a time I was invited to the High Table 
in an Oxbridge College and beside me there was this old physics professor. 
He was a nice man with lots of fun stories and I remember one of them 
in particular. Once he had done experiments on $10$ $MeV$ protons and 
he and his collaborators then saw large polarisation effects. But he 
said:` ``When I told the theorists of that time of the results 
they told me that if I would be able to perform the experiments 
with $20$ $MeV$ beams then these effects would go away!'' Now it is a 
fact that we have gone up to energies many thousands of times larger 
and we have looked all over and everywhere we look there are the same 
very big polarisation effects! So why have theorists so often claimed that 
such effects should go away? Because there is among many theorists the 
hope for simplicity so 
 that there should be a single channel amplitude that will dominate 
in Asymptopia.  
In most theoretical work you find that polarisation effects stem from 
interferences between different amplitudes and therefore they ``should'' 
go away.

A phenomenological string dynamics person like me would say that a 
confined force field always ``ought to'' produce polarisation. In this 
case there should be {\em a well-defined field direction},  $\vec{n}$,  
defined e.g. from the 
$3$ to the $\bar{3}$ charge along the field. Therefore for every particle 
with a momentum $\vec{p}$ there is a an axial vector $ \vec{A} \equiv 
\vec{n}\times \vec{p}$ that may together with the spin vector $\vec{S}$ 
be used to construct a scalar term in the Hamiltonian $\vec{S}\cdot\vec{A}$.
We should consequently expect transverse polarisation effects, i.e.
out of the plane spanned by the particle momentum and the field direction. 

Together with Gunnar Ingelman we presented in a Physics Letter from 1979
a simple model containing these features and we were succesful in 
describing the large $\Lambda$ particle polarisations seen in proton 
fragmentation regions. Consider the breakup of a stringlike
force field with a (constant) energy density $\kappa \simeq 1$ $GeV/fm$ into 
a $(q \bar{q})$-pair with transverse momentum $\pm \vec{k}_{\perp}$. In this 
way you have conserved momentum at the breakup. If the 
parton restmass is $\mu$ then you need in order to conserve the energy 
to produce the pair at a distance $2 \ell = 2 \mu_{\perp}/\kappa$ where 
$\mu_{\perp}= \sqrt{\mu^2 + \vec{k}_{\perp}^2}$. But then the orbital 
angular momentum is not conserved. It is easy to see that you obtain a
vector $\vec{L}$ pointing out of the plane with a size $L \simeq 2l k_{\perp}=
2\mu_{\perp}k_{\perp}/\kappa$, i.e. it is of the order of unity for an average 
transverse momentum of size $\simeq 0.3$ $GeV$. You may then conserve 
the total angular momentum $\vec{J}=\vec{L}+\vec{S}$ by polarising 
the $(q \bar{q})$-pair out of the plane oppositely to $\vec{L}$!  
In this way I have presented you with a production mechanism for the pair 
corresponding to (in spectroscopic notation) a $3P_0$ assignment, i.e.
the pair is produced with vacuum quantum numbers!

Some of the results shown by Dr Liang can be interpreted along these lines.
Let me present you with a kind of favorite experiment (some of my experimental
friends, S.O. Holmgren et al.  did something similar already in the end of 
the 70's).
Suppose that you would trigger on a $\Lambda$ and a strange vector meson 
in its neighbourhood, a $K^*$ or a $\Phi$. The $\Lambda$ is from an 
$SU(6)$ point of view a simple object, essentially made up of an 
$s$-quark and a $(ud)_0$, i.e. a combination of an up- and 
down-quark with vanishing spin and isospin. Consequently polarisation 
of a $\Lambda$ means polarisation of the $s$-quark!
Further the $\Lambda$-particle decays weakly and consequently reveals 
(and this is  a large effect) its spin in the decay distribution. 
The vector meson decays via strong interactions, which do not 
differ between up and down but you can make use of the $\Lambda$ 
decay to define the direction out of the plane. And my friends 
found very large transversity amplitudes in their studies. You should note
that while the string field spanned between the charges provide longitudinal
dynamics and the transverse momentum fluctuations in the breakup give 
the dynamics in one of the remaining space directions the polarisation 
as used in such an experiments will provide also information in the 
third direction!

We have also seen the (first) results of the RHIC experiments in all 
their glory! And we are amazed, both about the fact that the accelerator 
works so well and that the detectors are all up and running. And this is 
of course the most amazing: they are able to measure details 
in events containing 
thousands of central tracks! We note that RHIC is the first instrument 
able to produce a real central region in heavy ion physics, 
i.e. the rapidity range is 
large enough so that there is more than just the fragmentation regions 
from the target and the beam. Up to now it seems that the signals 
correspond to ``simple'' scaling up of the results from the SPS heavy 
ion program and there is no hints of a phase transition as of yet.

A general comment is that they see so much more strangeness than we 
have found in ordinary hadronic events (not to mention in  the ``ordinary''
$e^+e^-$-annihilation events with their rather low particle density). 
One way to explain it is to claim that in thermal equilibrium at 
a sufficiently high temperature we expect that the strangeness 
degrees of freedom are filled just as well as the ``ordinary'' up and 
down ones. Another (which may basically be the same) is 
to claim that in a dense 
hadronic gas it is an effect of rescatterings. Actually we have from 
the wonderful data set at the $Z^0$ pole such a large statistical sample of
events in $e^+e^-$-annihilation that we may study  particle densities 
essentially above the ordinary ones. I 
have asked Klaus Hamacher to investigate whether the strangeness content 
increases with the multiplicity. There is at least in the Lund Model no 
way to obtain high energy densities along the fields (particle production in
a Lund Model jet is still a few  particles per rapidity unit if you use 
a rapidity variable along the jet). 

Another generality is that all theoretical signals proposed for a quark-gluon 
plasma seem to be negative, i.e. there should be less $J/\Psi$ 
because of Debye screening, there 
should be less  energetic jets because of rescattering of the gluons 
on the way out etc. I, for one, would feel happy to hear at least one 
positive signal  proposal because what do we need a new state of matter to if 
it only contains negative features w.r.t. everything else?

Pino Marchesini introduced the subject of hard QCD. He told us on 
the one hand that nowadays there exist in some cases  exact perturbative 
calculations up to the order $\alpha_s^5$ (!). On the other hand 
the resulting comparisons with data lead in general to good results 
(at least after some help from the friends, if I may be impertinent-
``little help'' means adding some things  and subtracting 
some others, I will come back to that later). 

Pino, who is an honest gentleman 
also pointed to some places where there are clear deviations between the 
experimental findings and our present understanding of QCD. In particular 
he pointed to the so-called pedestal effect, i.e. the fact that there is 
much more activity in the rapidity neighbourhood of a large jet than in 
``minimum-bias events''. Including all the bremsstrahlung 
radiation from the in- and outgoing
partons participating in a hard Rutherford scattering one can raise 
the background level a bit but there is still a factor of more than 
two lacking in the particle production. Consequently it is necessary 
to introduce, besides the hard Rutherfords something new and extra. 
Sj\"ostrand in PYTHIA has introduced multiple parton interactions but
in order to raise the signal sufficiently much 
he had to introduce a kind of ``bunching'' of the partons so that if there 
is one hard interaction there are in general several. Pino briefly 
described another effort by himself and 
Bryan Webber inside the HERWIG scenario, to introduce ``beam-line radiation''. 
Actually I remember that inside the very simple FRITIOF scenarium 
which we introduced a long time ago (and which admittedly did not 
contain all the sophistication of today's QCD) there was no difficulty 
to obtain the pedestal.

Dr Behnke pointed to possible problems in heavy quark production in DIS.
Firstly it seems that the present signal for beauty exceeds the 
next-to-leading order calculations by a factor of three. Secondly he needed 
contributions both from ``ordinary'' boson-gluon fussion and from ``resolved''
photons (the latter was large) to describe the charm content. The 
difference between the two mechanisms introduces different 
correlations between the $c$ and the $\bar{c}$ partons. I 
understand that the new experiments in the upgraded HERA2 will be able 
to study this problem.

Let me also mention a careful study of how to differ between quark- and 
gluon jets by Dr Yu Meiling and the (expected) findings reported by 
Dr Heaphy that a quark jet is slimmer than a gluon jet in accordance 
with the models. Finally let us note that $\alpha_s$ is running all over 
the place and that we  saw a particularly beautiful picture of it 
from Dr Flagmeyer.

We were introduced in a nice way to our `` oldest participant'', 
i.e. the session on Correlations and Fluctuations by Wolfram Kittel. 
Nick var Remortel and Wes Metzger discussed different 
features of Bose-Einstein correlations. In particular 
the correlation pattern of charged pions indicates a ``radius'' of the 
emission region close to $1$ $fm$ independently of the reaction 
(besides heavy ion reactions), while the neutral pions seem to 
stem from a region that is $1-2$ $\sigma$ deviations smaller. Further 
one finds a shorter transverse than longitudinal size in the 
two-particle correlations.  Wes  also 
pointed out that the true three-particle correlations among identical bosons
were entirely given by a combination of the two-particle correlations
(at least if one uses the variable $Q_{123}^2 = Q_{12}^2 +Q_{23}^2+Q_{13}^2$
with $Q_{ij}^2= - (p_i -p_j)^2$. On a direct question he was reluctant to 
provide future partitioning of the data into longitudinal and transverse
directions and I understand that these measurements puts a very large strain
on the person doing them). In the conventional ``chaotic sources'' 
description of the correlations there is a possible angle between 
the two-particle Fourier transform of the sources and the Nijmegen 
result is that this angle vanishes.
 
A more puzzling result was discussed by van Dalen, i.e. the fact that 
there are no traces of Bose-Einstein correlations between the two $W$'s 
at LEPII (although all the experiments now show the same signals for the 
correlations inside each $W$ as they show on the $Z^0$-pole in LEPI).  
A very large amount of work has been done in this connection because of 
the possibility pointed out in some theoretical studies that the 
correlations may make the width of the $W$ uncertain by possibly much 
more than the hoped-for $40$ $MeV$ (needed to constrain the Higgs mass).
Although the model I have myself built (together with Werner Hoffman (1986)
and Marcus Rign\'er (1996-98)) does not provide for such 
``inter-correlations'' it is evident that the conventional model should 
expect them. The reason for the lack of inter-correlations may actually 
be that LEPII has been too succesful in going up in energies 
(which some people have been pushing for very strongly in the hope 
of finding the Higgs). It is evident that it is the slow particles produced 
``in-between'' the two $W$'s that would have been able to show an effect.
The larger the energy produced in LEPII the further the $W$'s will move 
away from each other and therefore the smaller would be the effect. 

We also heard in a very interesting aside a report on Very Long 
Baseline Interferometry as it is used in astronomy (I understand from 
Dr Gurvits that the astromers have come along way from Hanbury-Brown and 
Twiss!).

I will  from the last day (my only excuse is that I was pretty 
tired after having 
written my transparences the night before) only mention one very 
interesting parametrisation for the structure functions 
provided in the talk by David Milstead. It has been known for a long time that
the DGLAP mechanism (corresponding to emission chains with well-ordered 
virtualities-transverse momenta along the ladders) does a very good job 
to describe the structure functions.A typical behaviour stemming from the 
summing up of such chains would be $F_2 \simeq \exp(C \sqrt{log(1/x)}$) with
$x$ the longitudinal scaling variable and $C$ a slowly varying function 
of the measured virtuality $Q^2$). The competing BFKL mechanism 
(corresponding to a possible going up and down in virtuality) would result in 
a typical behaviour like $F_2 \simeq D(Q^2) x^{- \lambda}$ with 
$\lambda$ the eigenvalue of a complicated equation, but (after some 
corrections)expected to be $\simeq 0.3-0.4$. Such a behaviour has been 
looked for extensively because of its possible relationship to the 
elusive ``hard funny-P'' or whatever people call the instigator of 
diffraction. Needless to say the BFKL mechanism has not been pinned down
(although the jury is out still). One reason is that despite its 
wonderful performance the range of $x$-values accesible in HERA may be
much too small. An estimate on the emission rate of gluons along the ladders
would tell us that there are only two to four emissions available between
$x\simeq 10^{-2}$ to $x\simeq 10^{-5}$. We can hardly expect that to 
be sufficient to drive the equations to a steady state (Gustafson provided a
pedagogical way to see the emergence of BFKL in a simplified scenarium).

Nevertheless David provided in his last transparency a parametrisation
of $F_2$ as $F_2 \simeq D \exp(-a \log(x)\log(Q^2/Q_0^2))$ with $D$ 
a constant $\simeq 
0.18$, $a \simeq 0.048$ and $Q_0 (\simeq \Lambda) \simeq 0.26$ $GeV$. 
This would mean that the experimentalists are seeing a BFKL shape but 
with a ``running'' $\lambda$-value, i.e. it looks as if $\lambda \simeq 
E/\alpha_s(Q^2)$ with $E$ a constant! 
I should immediately say that David only presented the data
and pointed out that the $\lambda$ parameter was a linear function 
of $\log(Q^2)$. But I cannot help but to take it all the way 
(although the data shown were  only for $Q>1.5$ $GeV$)! Whatever the 
cause for the remarkably simple parametrisation it is not the ``ordinary'' 
BFKL because in that case the $\lambda$ parameter is proportional to
$\alpha_s$.

\section{Things Dear to My Heart}

\subsection{A Lamentation and A Scale for The Border region}

QCD is doubtlessly the greatest intellectual challenge and adventure 
of my generation of physicists and it is so different from the field 
theories we discussed in Lund in the $60$'s! The gauge field self-interaction 
implies e.g. that the imaginary part of the polarisation tensor changes sign
compared to anything we had seen when we summed over a positive definite 
metric for the asymptotic states. (This is a complex way to say that the 
running coupling vanishes at large $Q^2$ but at least older people may
understand the surprise). The perturbation theory that Feynman and 
his contemporaries provided us with certainly works in QED with its small 
effective coupling and the vacuum fluctuations can (after renormalisation)
be handled (and even understood!) just as small deviations from 
the no-particle state. But we 
know from the very beginning that the corresponding treatment of QCD must
lead to all kinds of problems and they certainly do! For every quantity 
that is calculated we need to introduce ``resummations of leading or 
subleading quantities''. I am impressed about the technical skill that has 
been developed but I also have a feeling that the physics is slipping away
from us. Maybe we are in the future only going to hear signals every now
and then from this community in the same way as we now do from the lattice 
people.

The problems that we are facing now are, however, inside the border region
between where present day perturbation theory (with all its extras) is expected
to work and the region where we truly know that non-perturbative models are 
necessary. So let us start to try to define it. Inside the Lund Model (and
frankly I expect that the results are pretty general) there are two scales 
that seem to be the same. The first is the scale where we can at first hope 
to be able to disentangle ``real'' gluons, i.e. objects that will stick 
out of the confining force field. The second scale is the size of the momentum 
transfers between the hadrons produced in the fragmentation process. 

To see the first scale I will make use of an argument similar to the 
Landau-Pomeranchuk formation time. In that case the question is: ``when can 
we differ between the state of a single (charged) electron and a 
state containing the electron and a photon?'' They argued that if we assume 
that the electron moves along the z-axis and the photon has energy $e$ and 
tranverse momentum $k_{\perp}$ (for definiteness along the x-axis) then 
we can boost to a frame where the photon only has momentum along the 
x-axis. In that frame the formation time must correspond to  at least 
a wave-length $\tau \simeq 2\pi/k_{\perp}$ because you cannot ``see'' 
a wave before that.
In the earlier frame that corresponds to the ``formation time'' 
$2\pi e/k_{\perp}^2$, and with the $k_{\perp}$ exchanged for some suitable 
``virtuality scale'' it works as an estimate of what quantum mechanics 
and relativity means in a general case (e.g. in the Lund Model it is always 
the slowest particles that are first disentangled in the ``inside-out'' 
cascade'').

To see the confinement scale suppose that we consider a gluon moving 
transverse to the force field with energy and momentum equal to $k_{\perp}$. 
Then the gluon can move out at most the length $\ell = k_{\perp}/2\kappa$
(the force on the gluon is in the Lund Model twice the string constant 
$\kappa$). In order to be ``real'' this must correspond at least to a 
wave length, i.e. $\ell \geq 2\pi/k_{\perp}$ or $k_{\perp}^2 \geq 4\pi \kappa
\simeq 2.5$ $GeV^2$. (The width of the string field should be $\ell \simeq 
\sqrt{\pi/\kappa}$ and this is also obtained from the ``ordinary'' 
tunneling arguments). At first sight this seem to be a large scale but 
it is interesting to note that it is the same scale as will occur for the 
momentum transfers between the particles in the fragmentation process (and
it is to compare to something else twice the (inverse) 
slope of the Regge trajectories).

The Lund Fragmentation Model is based upon the Area Law, i.e the 
(non-normalised) probability to produce $N$ particles with the energy-momenta
$\{p_j\}$ given the total energy momentum $P_{tot}$ is 
\begin{eqnarray}
\label{arealaw}
dP_N = \prod_{j=1}^N N_j dp_j \delta(p_j^2-m_j^2) \delta(\sum p_j-P_{tot}) 
\exp(-b A)
\end{eqnarray}
with $N$ a normalisation constant and $A$ the area of the string before 
it decays. We have often interpreted this in accordance with Fermi's 
Golden Rule as the phase space of the final state particles multiplied by
the squared transition matrix element. In a paper (hep-ph/9910374) Fredrik
S\"oderberg and I reinterpreted the result in Eq. (\ref{arealaw}) as a 
product of transition operators (in quantum mechanics it would be 
density operators) one for each particle between the momentum transfer states 
$\{q_j\}$ such that $p_j = q_{j-1}-q_{j}$. We found that the operators can 
be diagonalised as
\begin{eqnarray}
\label{operdiag}
<q_j|O|q_{j-1}> = \sum_n g_n(b\Gamma_j)\lambda_n(bm_j^2)g_n(b\Gamma_{j-1})
\end{eqnarray}
in terms of the two-dimensional harmonic oscillator eigenfunctions 
($\Gamma \equiv -q^2$)  and with
the eigenvalues $\lambda_n$ analytical continuations of the $g_n$ into 
time-like arguments ($p_j^2=m_j^2)$. 

Just for fun let me point out that (almost) 
everything you can do with the (free) 
plain-wave momentum eigenstates $\exp(ikx)$ you can also do with 
the harmonic oscillator wave functions. As an 
example, Fredrik and I proved 
that the total integrals of the $dP_N$ over the $N$-particle phase space
\begin{eqnarray}
\label{RNint}
\int dP_n = R_N(s) & \mbox{with} & s=P_{tot}^2
\end{eqnarray}
can be expanded in a simple way
\begin{eqnarray}
\label{Rint}
\int \frac{ds R_N(s)}{s+u} = \sum \lambda^N L_n(bu)
\end{eqnarray}
in terms of the (two-dimensional) harmonic oscillator correspondences 
to the Hermite polynomials, the Laguerre polynomials. You may remember the 
nice lecture by Engel on the Sommerfeld-Watson transform done 
on the partial wave (this corresponds to Laplace polynomials)
sums of the elastic amplitude and its relations to Regge trajectories. You can
do exactly the same transform on the sum over all the multiplicities of $R_N$
in Eq. (\ref{RNint}), $\sum R_N=R$ by means of the sum in Eq. (\ref{Rint})
to obtain that $R(s) \simeq s^a$ where the parameter $a$ is the 
correspondence to a Regge intercept in the Lund Model. 

The parameter 
$a= a(N, bm^2)$ is phenomenologically close to $0.5$ and it also regulates 
the behaviour of the average momentum transfer, $\Gamma$, size in the 
Lund Model. The inclusive $\Gamma$ behaviour is 
$\propto \Gamma^a \exp(-b\Gamma)$ which implies  an average size $<\Gamma>=
(1+a)/b \simeq 2.5$ $GeV$, i.e. the same size as the minimum size of 
the ``real'' gluons defined above! Maybe the occurrence of the harmonic 
oscillator wave functions, that evidently contains information of confinement 
in the string field or some similar functions should play a major role in the 
future work on confinement in QCD?

\subsection{On Some Further Developments Based on the Area Law}

I have in earlier talks at these conferences pointed out that if we interprete
the Area Law as the phase space times the square of a transition matrix 
element then there is an obvious candidate for such a matrix element, i.e. 
it is  a Wilson loop operator evaluated over the confined string 
field during its breakup. The reason is that the particles in the Lund Model
are produced over a region of the field, i.e. a (color-singlet) hadron stems
from a quark from one breakup vertex and an anti-quark from an adjacent 
vertex and the field in between. In order to keep to gauge invariance it 
is then necessary to have a gauge connector between the vertices 
$\exp(i \int gA^{\mu}dx_{\mu})$. 

Using the same argument for all the adjacent vertices we find that  
the whole state must be endowed with
$\exp(i \oint gA^{\mu}dx_{\mu}) = \exp(i\xi A)$, where the Wilson loop 
goes over the field region with the (breakup) area $A$. The parameter $\xi$ 
has a real part given by the string constant (this is used in lattice 
gauge calculations). In a decay situation $\xi$ must  also have an 
imaginary part corresponding to ``absorption''. In the Kramers-Kronig
interpretation the imaginary part of the dielectricity is related to the pair 
production rate. I have in my book (``The Lund Model'') presented detailed 
calculations and I feel that the size of the Lund Model parameter $b$ in the 
Area Law $b\simeq 0.6$ $GeV^{-2}$ fits well inside that picture.

If this is taken seriously then we have a model with a matrix element 
containing a complex phase. Such a phase will be noticable e.g. if 
there are two identical bosonic particles produced in the state because
then the matrix element  must be symmetrical in the particle variables.
It turns out (in the simple case when the string state stems from an original 
$(q \bar{q})$ pair and there are no gluonic excitations) that the same 
state can be produced if the two particles are exchanged 
in the production process with all the rest of the state unchanged. 
We then obtain in 
an easily understood notation the total matrix element 
${\cal M}\equiv {\cal M}_{12} + {\cal M}_{21}$. The square of the 
matrix element ${\cal M}$
will contain both the possibility to make the production in the order 
$(12)$ or in the order $(21)$ (with everything else the same). But there 
will also be an interference term $ I=\cos (\Delta A)/\cosh(B \Delta A + 
C\delta(p_{\perp}^2))$ with $\Delta A$ the area difference between the 
two configurations and $\delta(p_{\perp}^2)$ the necessary change in the 
$p_{\perp}$ generation according to  the tunneling mechanism in 
the Lund Model. Further 
$B$ is a (small) and $C$ an essentially larger parameter. 

You should note that $\Delta A$ is solely an energy momentum space quantity 
in this interpretation. It is given by $\Delta A = (p_1-p_2)\cdot \Delta$
with $\Delta$ the (space-like) energy-momentum content in the string produced 
between the  two identical particles with energy-momenta $p_1$ and $p_2$ 
(counted in inverse  string constant units). It is evident that 
the interference pattern is in this model sensitive to 
the energy momentum region inside which the quantum numbers   
of the particles are locally compensated.

The mechanism has been termed ``the string symmetrisation (or coherence) 
scheme'' by Eddi de Wolfe and the resulting formulas has  similarities 
(albeit it is a
very different dynamical mechanism) to the ordinary scheme where the basic 
assumption is that the production region is completely chaotic or incoherent.
It is, however, well-known that the Low theorem seem to work for the 
description of photon emission in the wavelength region used for the study 
of the interference pattern for the identical hadrons. Low's theorem is 
certainly based upon coherence and as I have said it before there are 
no reasons to believe that there are incoherent sources 
inside a multiparticle production region similar to those 
occurring for photon emission from a large star surface. The model 
describes the properties mentioned 
in the session on Correlations and Fluctuations (the fact that the neutral 
pions (may) have smaller size parameter than the charged is because they 
can in the Lund Model be produced adjacent in rank, the fact that 
longitudinal and transverse sizes are different is due to the size of 
the governing parameters etc.) I would, however, like to go on and 
briefly describe
what we learned from a recent study of the multigluon states.

I have (together with two great graduate students, Sandipan Mohanty 
and Fredrik S\"oderberg) written one paper (hep-ph/0106185) on the 
Lund Area Law for multigluon states and when 
I write this there are more to appear, hopefully within a month or two.
The work has been done because the well-known JETSET Monte Carlo will for 
multigluon states produce particles with only an approximate implementation 
of the Lund Area Law, i.e. inclusively it works very well but to study 
string coherence and other structure we evidently need the precise area. 

The final results can be briefly described for 
$e^+e^-$-annihilation events: 

\begin{itemize}

\item[A] Given a partonic state (e.g. from a perturbative cascade) with the 
partonic energy-momentum vectors $k_1,k_2 \ldots k_n$ in color order then 
we can construct a four-vector valued curve, the directrix curve 
${\cal A}_{\mu}$ by laying them out in order. The string surface used 
in the Lund Model as a model for the QCD force fields is a minimal
surface and it is consequently completely described by its boundary 
curve which turns out 
to be just this directrix curve. Therefore the breakup of the string can be
just as well described ``along the directrix curve'' as ``on the string 
surface''. 

\item[B] The fragmentation process correponds to the production of 
another curve, the $X$-curve, with the hadron energy momenta $\{p_j\}$ 
laid out in rank-order. The relationship between the ${\cal A}$ and the 
$X$-curves can be described as the formation of an area in between 
them, composed out of four-cornered ``plaquettes''. Each plaquette 
is bordered by a particle momentum (along the $X$-curve), a 
piece of the directrix $\delta{\cal A}$ and two (time-like) vectors $x$ 
such that
\begin{eqnarray}
\label{plaquet}
x_{j-1}+\delta{\cal A}_j = x_j +p_j
\end{eqnarray}
The area
of each plaquette correspond to the (sub)areas related to each particle 
production (as in the transfer operators mentioned in connection with 
Eq.(\ref{operdiag})). The length of the vectors $x_j$ fulfil $x_j^2=\Gamma_j$
in that equation).
 
\end{itemize} 

Although this description looks very abstract you can intuitively 
consider it so 
that each particle obtains its energy momentum both 
from a ``new'' part of the directrix, $\delta{\cal A}_j$ and also 
from
the ``remaining'' energy momenta of the ``earlier'' parts of the directrix 
through the $x_{j-1}$. The ``new'' remainder is then brought forward to 
the next production by the vertex vector $x_j$.

\subsection{Further Developments Based upon the Structure of the Parton 
Cascades and the Fragmentation Process}
There is no time to cover the many delightful properties we obtain but 
I would like to mention two further features of the model.
\begin{itemize}
\item [C] There is a limiting situation correponding to a vanishing 
mass-value. In that case the $X$-curve goes over to a continuous curve, 
the ${\cal X}$-curve that is characteristical of the particular partonic 
state but also in a very precise way corresponds to the ``average'' 
hadronic $X$-curve obtained from the stochastical fragmentation process.
\item[D] The vectors $x$ goes over to the time-like tangents of 
the ${\cal X}$-curve reaching to the directrix. They quickly 
approach a constant length, $m_0$. The ${\cal X}$-curve in this way 
looks like a set of connected hyperbolas, spanned on the ``distance'' $m_0$ 
between the lightlike parton energy momenta in the directrix. In this way 
$m_0$ may be considered as a ``resolution scale'' for the partonic 
directrix curve. The area between the directrix and the ${\cal X}$-curve 
is given by $m_0^2 \lambda(m_0)$.  
\end{itemize}
This generalised rapidity $\lambda(m_0)$ that comes out of the Lund Area Law
fragmentation process also comes out of (at least one of) the perturbative 
parton cascades, viz. the Lund Dipole Cascade Model as 
it is implemented in the Monte Carlo program ARIADNE! (I believe that 
there are correspondences in  both HERWIG and JETSET). As I have talked 
about this model at earlier meetings I will be very brief.

We firstly note the two basic formulas for QCD bremsstrahlung. The dipole
bremsstrahlung formula for the inclusive production of gluons is
\begin{eqnarray}
\label{brems1}
dn= \alpha_{eff} dy \frac{dk_{\perp}^2}{k_{\perp}^2} (Pol-sum)
\end{eqnarray}
(where $(Pol-sum)$ is the coupling of the spins of the emitters and the gluon.
It is close to unity except for collinear gluon emissions).
The emission of two gluons from an original $(q \bar{q})$-dipole is 
factorisable and can be written (besides a small correction)
\begin{eqnarray}
\label{brems2}
dn(qg_1g_2\bar{q})= dn(qg_1\bar{q})(dn(qg_2g_1)+dn(g_1g_2\bar{q}))
\end{eqnarray}
Eq.(\ref{brems2}) is valid if $k_{\perp 1}>k_{\perp 2}$ or else the 
two gluons are exchanged. Note that the two ``new'' dipoles'' are moving 
w.r.t. each other (and also remember the remarks in the Introduction). 
The coherence conditions limits the phase space 
(in this case just as energy momentum conservation would do!) so that in the 
the dipole restframe (with the dipole squred mass equal to $s$) we obtain
\begin{eqnarray}
\label{energymomcons}
k_{\perp} \cosh(y) \leq \frac{\sqrt{s}}{2}
\end{eqnarray}
i.e. essentially the region $|y| < 1/2\log(s/k_{\perp}^2)$ 
so that the rapidity range for a dipole emission is 
$\Delta y \simeq \log(s/k_{\perp}^2)$. If we denote 
the three partons after the first emission $(1,3)$ (the emitters) and 
$2$ (the gluon) then the combined rapidity range for a second emission 
either from the dipole $(12)$ or from the dipole $(23)$ is
\begin{eqnarray}
\label{raprange1}
(\Delta y)_{gen}= (\Delta y)_{12} + (\Delta y)_{23} = \\
\log (s_{12}/2k_{\perp}^2) + 
\log(s_{23}/2k_{\perp}^2)=\log(s/k_{\perp}^2) + 
\log(s_{12}s_{23}/4sk_{\perp}^2)
\end{eqnarray}
The argument in the last logarithm is $s_{12}s_{23}/s=k_{\perp 2}^2$, i.e.
an invariant definition of the first gluon transverse momentum squared. 
In this way the first emission will increase the rapidity range for the second
(the variable $s$ is still the total cms mass squared. The factor $1/2$ 
in each of the dipoles  is due to the fact that only half of the first gluon
goes into each dipole). If the first emission is ``soft'' or ``collinear'' 
then there is a correction to the formula so that we obtain
\begin{eqnarray}
\label{raprange2}
\lambda(m_0^2) \cong \log(s/m_0^2 + s_{12}s_{23}/4m_0^4)
\end{eqnarray}
i.e. a nice interpolation between the emission from one and from two dipoles.
(the sign $\cong$ is used because we will for factorisation purposes add
a constant in the argument of the logarithm). This is then the 
$\lambda$-measure 
for a state with one gluon and it can be extended in an infrared safe way 
to the multigluon states. Then it coincides with the results from the 
length of the ${\cal X}$-curve that we obtained in the fragmentation process 
as I described it above.

I will end with a further very remarkable property of the $\lambda$-measure.
The Dipole Cascade Model is built in such a way that you start with a single
dipole and then you ``go downwards'' in transverse momentum until 
it breaks up into two dipoles (this means in technical language that 
there is a Sudakov form factor). Then you continue downwards all the time 
looking for new emssions thereby producing new dipoles. In that way the 
transverse momentum at every stage is both an ordering and a resolution 
parameter. In ARIADNE the process is implemented with a running 
$\alpha_{eff}$, a precise implementation of the $(Pol-sum)$ and  finally 
with a local energy-momentum conservation so that the emitting dipole takes 
the recoil. A hard gluon emission will increase the phase space, i.e. the 
size of the $\lambda$-measure, thereby  opening  up for more 
and more softer gluon emissions. In this way the dipoles are  spreading 
and moving away in different directions, although in the ``ordinary'' 
phase space with the rapidity e.g. defined along some thrust axis 
they will seem to be collimated along a set of jet directions.It 
is evidently interesting to look for the resulting ``local'' properties 
of the cascade, i.e. to look for the distribution of gluons (or equivalently
dipoles) along the color flow, i.e. along the ${\cal X}$-curve and the 
corresponding $\lambda$-measure.

Due to the nice properties of the $\lambda(m_0)$ function it is possible 
at every scale to define it as an independent sum of contributions one 
for each gluon that is resolved on the scale $m_0$:
\begin{eqnarray}
\label{deltalambda}
\lambda(m_0)= \sum_{j=1}^n(\Delta \lambda)_j(m_0) & \mbox{with} & 
(\Delta \lambda)_j=\log(1+x_{j-1}k_j/m_0^2)
\end{eqnarray}
The great surprise is that the distribution in $(\Delta \lambda)$ depends 
rather little upon the scale $m_0$ and it is the same for all events 
independently of thrust, sphericity or other global ``excitation'' variables!
It has an average value around three units and a width around one and looks
very much like the well-known mathematical $\Gamma$-distribution, i.e. it has
an exponential tail. Remembering the relationship between the $\lambda$-measure
and an area it seems as if we are getting back a new area law, this time for
the parton cascade! We are evidently at the moment investigating these 
properties in detail and as I write this I know much more than I knew when 
I gave the talk. But besides one feature this will be for another occassion.

The only feature I would like to mention is that this partitioning of the 
$\lambda$-measure  into $(\Delta \lambda)$ pieces has a 
direct reflection into the 
properties of the fragmentation process. Each piece (which you could call
a generalised dipole region) decays independently of the others into a set of
hadrons that are all in an essentially ``planar'' state, i.e. besides (small) 
transverse momentum fluctuations they all lie in a $(1+1)$-dimensional 
(time-like) subspace. This is just as in the original Lund Fragmentation
Model stemming from a $(q \bar{q})$-state with no internal excitations.
Therefore we can in each generalised dipole region do string symmetrisation 
in accordance with
the model for Bose-Einstein correlations. You can look upon these regions
as ``coherent sources''. It seems to occur very seldomly that particles 
stemming from different such sources are so close in energy momentum that 
they should be able to interfere. This work is not finished yet but as it 
seems at the moment string coherence should give the right correlation pattern
also for the multigluon states.

\section*{Acknowledgments}
I would like to thank the organisers both for a very nice meeting and for
asking me to give this closing talk.

\end{document}